\documentclass[useAMS,usenatbib]{mn2e}
\usepackage{tabularx}
\usepackage{xspace}
\usepackage{graphicx}
\newcommand{\ignore}[1]{}
\providecommand{\ao}{}

\renewcommand{\ao}{adaptive optics (AO)\renewcommand{\ao}{AO\xspace}\renewcommand{\Ao}{AO\xspace}\xspace}
\newcommand{\Ao}{Adaptive optics (AO)\renewcommand{\ao}{AO\xspace}\renewcommand{\Ao}{AO\xspace}\xspace}
\newcommand{\wfs}{wavefront sensor (WFS)\renewcommand{\wfs}{WFS\xspace}\renewcommand{\wfss}{WFSs\xspace}\xspace}
\newcommand{\wfss}{wavefront sensors (WFSs)\renewcommand{\wfs}{WFS\xspace}\renewcommand{\wfss}{WFSs\xspace}\xspace}
\newcommand{\shwfs}{Shack-Hartmann \wfs (SHWFS)\renewcommand{\shwfs}{SHWFS\xspace}\xspace}
\newcommand{\dm}{deformable mirror (DM)\renewcommand{\dm}{DM\xspace}\renewcommand{\dms}{DMs\xspace}\renewcommand{\Dms}{DMs\xspace}\renewcommand{\Dm}{DM\xspace}\xspace}
\newcommand{\dms}{deformable mirrors (DMs)\renewcommand{\dm}{DM\xspace}\renewcommand{\dms}{DMs\xspace}\renewcommand{\Dms}{DMs\xspace}\renewcommand{\Dm}{DM\xspace}\xspace}
\newcommand{\Dms}{Deformable mirrors (DMs)\renewcommand{\dm}{DM\xspace}\renewcommand{\dms}{DMs\xspace}\renewcommand{\Dms}{DMs\xspace}\renewcommand{\Dm}{DM\xspace}\xspace}
\newcommand{\Dm}{Deformable mirror (DM)\renewcommand{\dm}{DM\xspace}\renewcommand{\dms}{DMs\xspace}\renewcommand{\Dms}{DMs\xspace}\renewcommand{\Dm}{DM\xspace}\xspace}

\newcommand{\lqg}{linear-quadratic-gaussian (LQG)\renewcommand{\lqg}{LQG\xspace}\xspace}
\newcommand{\shs}{Shack-Hartmann sensor (SHS)\renewcommand{\shs}{SHS\xspace}\renewcommand{\shss}{SHSs\xspace}\xspace}
\newcommand{\shss}{Shack-Hartmann sensors (SHSs)\renewcommand{\shs}{SHS\xspace}\renewcommand{\shss}{SHSs\xspace}\xspace}
\newcommand{\lgs}{laser guide star (LGS)\renewcommand{\lgs}{LGS\xspace}\renewcommand{\lgss}{LGSs\xspace}\xspace}
\newcommand{\lgss}{laser guide stars (LGSs)\renewcommand{\lgs}{LGS\xspace}\renewcommand{\lgss}{LGSs\xspace}\xspace}
\newcommand{\Ngs}{Natural guide star (NGS)\renewcommand{\ngs}{NGS\xspace}\renewcommand{\Ngs}{NGS\xspace}\renewcommand{\ngss}{NGSs\xspace}\xspace}
\newcommand{\ngs}{natural guide star (NGS)\renewcommand{\ngs}{NGS\xspace}\renewcommand{\Ngs}{NGS\xspace}\renewcommand{\ngss}{NGSs\xspace}\xspace}
\newcommand{\ngss}{natural guide stars (NGSs)\renewcommand{\ngs}{NGS\xspace}\renewcommand{\Ngs}{NGS\xspace}\renewcommand{\ngss}{NGSs\xspace}\xspace}
\newcommand{\mems}{Micro-Electro-Mechanical Systems (MEMS)\renewcommand{\mems}{MEMS\xspace}\xspace}
\newcommand{\snr}{signal to noise ratio (SNR)\renewcommand{\snr}{SNR\xspace}\xspace}
\newcommand{\Moao}{Multi-object \ao (MOAO)\renewcommand{\moao}{MOAO\xspace}\renewcommand{\Moao}{MOAO\xspace}\xspace}
\newcommand{\moao}{multi-object \ao (MOAO)\renewcommand{\moao}{MOAO\xspace}\renewcommand{\Moao}{MOAO\xspace}\xspace}
\newcommand{\mcao}{multi-conjugate adaptive optics (MCAO)\renewcommand{\mcao}{MCAO\xspace}\xspace}
\newcommand{\ltao}{laser tomographic \ao (LTAO)\renewcommand{\ltao}{LTAO\xspace}\xspace}
\newcommand{\cpu}{central processing unit (CPU)\renewcommand{\cpu}{CPU\xspace}\renewcommand{\cpus}{CPUs\xspace}\xspace}
\newcommand{\cpus}{central processing units (CPUs)\renewcommand{\cpu}{CPU\xspace}\renewcommand{\cpus}{CPUs\xspace}\xspace}
\newcommand{\psf}{point spread function (PSF)\renewcommand{\psf}{PSF\xspace}\renewcommand{\psfs}{PSFs\xspace}\renewcommand{\Psf}{PSF\xspace}\xspace}
\newcommand{\psfs}{point spread functions (PSFs)\renewcommand{\psf}{PSF\xspace}\renewcommand{\psfs}{PSFs\xspace}\renewcommand{\Psf}{PSF\xspace}\xspace}
\newcommand{\Psf}{Point spread function (PSF)\renewcommand{\psf}{PSF\xspace}\renewcommand{\psfs}{PSFs\xspace}\renewcommand{\Psf}{PSF\xspace}\xspace}
\newcommand{\fpga}{field programmable gate array (FPGA)\renewcommand{\fpga}{FPGA\xspace}\renewcommand{\fpgas}{FPGAs\xspace}\xspace}
\newcommand{\fpgas}{field programmable gate arrays (FPGAs)\renewcommand{\fpga}{FPGA\xspace}\renewcommand{\fpgas}{FPGAs\xspace}\xspace}
\newcommand{\sor}{successive over-relaxation (SOR)\renewcommand{\sor}{SOR\xspace}\xspace}
\newcommand{\fdpcg}{Fourier domain pre-conditioned gradient (FDPCG)\renewcommand{\fdpcg}{FDPCG\xspace}\xspace}
\newcommand{\map}{maximum a-posteriori (MAP)\renewcommand{\map}{MAP\xspace}\xspace}

\newcommand{\elt}{Extremely Large Telescope (ELT)\renewcommand{\elt}{ELT\xspace}\renewcommand{\elts}{ELTs\xspace}\renewcommand{\eelt}{European ELT (E-ELT)\renewcommand{\eelt}{E-ELT\xspace}\xspace}\xspace}

\newcommand{\elts}{Extremely Large Telescopes (ELTs)\renewcommand{\elt}{ELT\xspace}\renewcommand{\elts}{ELTs\xspace}\renewcommand{\eelt}{European ELT (E-ELT)\renewcommand{\eelt}{E-ELT\xspace}\xspace}\xspace}

\newcommand{\eelt}{European Extremely Large Telescope (E-ELT)\renewcommand{\eelt}{E-ELT\xspace}\renewcommand{\elt}{ELT\xspace}\renewcommand{\elts}{ELTs\xspace}\xspace}

\newcommand{\dugall}{Durham University generalised adaptive optics laser laboratory (DUGALL)\renewcommand{\dugall}{DUGALL\xspace}\xspace}
\newcommand{\fwhm}{full-width at half-maximum (FWHM)\renewcommand{\fwhm}{FWHM\xspace}\xspace}
\newcommand{\wht}{William Herschel Telescope (WHT)\renewcommand{\wht}{WHT\xspace}\xspace}
\newcommand{\emccd}{electron multiplying CCD (EMCCD)\renewcommand{\emccd}{EMCCD\xspace}\renewcommand{\emccds}{EMCCDs\xspace}\xspace}
\newcommand{\emccds}{electron multiplying CCDs (EMCCDs)\renewcommand{\emccd}{EMCCD\xspace}\renewcommand{\emccds}{EMCCDs\xspace}\xspace}
\newcommand{\dasp}{Durham \ao simulation platform (DASP)\renewcommand{\dasp}{DASP\xspace}\renewcommand{\thedasp}{DASP\xspace}\renewcommand{\Thedasp}{DASP\xspace}\xspace}
\newcommand{\thedasp}{the Durham \ao simulation platform (DASP)\renewcommand{\dasp}{DASP\xspace}\renewcommand{\thedasp}{DASP\xspace}\renewcommand{\Thedasp}{DASP\xspace}\xspace}
\newcommand{\Thedasp}{The Durham \ao simulation platform (DASP)\renewcommand{\dasp}{DASP\xspace}\renewcommand{\thedasp}{DASP\xspace}\renewcommand{\Thedasp}{DASP\xspace}\xspace}
\newcommand{\mpi}{Message Passing Interface (MPI)\renewcommand{\mpi}{MPI\xspace}\xspace}
\newcommand{\smp}{symmetric multi-processing (SMP)\renewcommand{\smp}{SMP\xspace}\xspace}
\newcommand{\svd}{singular value decomposition (SVD)\renewcommand{\svd}{SVD\xspace}\xspace}
\newcommand{\gpu}{graphics processing unit (GPU)\renewcommand{\gpu}{GPU\xspace}\renewcommand{\gpus}{GPUs\xspace}\xspace}
\newcommand{\gpus}{graphics processing units (GPUs)\renewcommand{\gpu}{GPU\xspace}\renewcommand{\gpus}{GPUs\xspace}\xspace}
\newcommand{\fft}{fast Fourier transform (FFT)\renewcommand{\fft}{FFT\xspace}\xspace}
\newcommand{\ifu}{integral field unit (IFU)\renewcommand{\ifu}{IFU\xspace}\xspace}
\newcommand{\darc}{the Durham \ao real-time controller (DARC)\renewcommand{\darc}{DARC\xspace}\renewcommand{\Darc}{DARC\xspace}\xspace}
\newcommand{\Darc}{The Durham \ao real-time controller (DARC)\renewcommand{\darc}{DARC\xspace}\renewcommand{\Darc}{DARC\xspace}\xspace}
\newcommand{\cots}{commercial off-the-shelf (COTS)\renewcommand{\cots}{COTS\xspace}\xspace}
\newcommand{\rtcp}{real-time control pipeline (RTCP)\renewcommand{\rtcp}{RTCP\xspace}\xspace}
\newcommand{\rms}{root-mean-square (RMS)\renewcommand{\rms}{RMS\xspace}\xspace}
\newcommand{\sFPDP}{serial Front Panel Data Port (sFPDP)\renewcommand{\sFPDP}{sFPDP\xspace}\xspace}
\newcommand{\wpu}{wavefront processing unit (WPU)\renewcommand{\wpu}{WPU\xspace}\xspace}

\newcommand{\rtcs}{real-time control system (RTCS)\renewcommand{\rtcs}{RTCS\xspace}\renewcommand{\rtcss}{RTCSs\xspace}\xspace}
\newcommand{\rtcss}{real-time control systems (RTCSs)\renewcommand{\rtcs}{RTCS\xspace}\renewcommand{\rtcss}{RTCSs\xspace}\xspace}
\newcommand{\eso}{European Southern Observatory (ESO)\renewcommand{\eso}{ESO\xspace}\renewcommand{\theeso}{ESO\xspace}\xspace}
\newcommand{\theeso}{\renewcommand{\theeso}{ESO\xspace}the \eso}
\newcommand{\scao}{single conjugate \ao (SCAO)\renewcommand{\scao}{SCAO\xspace}\renewcommand{\Scao}{SCAO\xspace}\xspace}
\newcommand{\Scao}{Single conjugate \ao (SCAO)\renewcommand{\scao}{SCAO\xspace}\renewcommand{\Scao}{SCAO\xspace}\xspace}
\newcommand{\glao}{ground layer \ao (GLAO)\renewcommand{\glao}{GLAO\xspace}\xspace}
\newcommand{\eagle}{ELT Adaptive optics for GaLaxy Evolution (EAGLE)\renewcommand{\eagle}{EAGLE\xspace}\xspace}
\newcommand{\maory}{multi-conjugate \ao relay for the \eelt (MAORY)\renewcommand{\maory}{MAORY\xspace}\xspace}
\newcommand{\muse}{Multi Unit Spectroscopic Explorer (MUSE)\renewcommand{\muse}{MUSE\xspace}\xspace}
\newcommand{\vlt}{Very Large Telescope (VLT)\renewcommand{\vlt}{VLT\xspace}\xspace}

\newcommand{\tmt}{Thirty Metre Telescope (TMT)\renewcommand{\tmt}{TMT\xspace}\xspace}
\newcommand{\xao}{eXtreme \ao (XAO)\renewcommand{\xao}{XAO\xspace}\xspace}

\newcommand{\vla}{Very Large Array (VLA)\renewcommand{\vla}{VLA\xspace}\xspace}
\newcommand{\jwst}{{\em James Webb Space Telescope} \citep[JWST,][]{jwst}\renewcommand{\jwst}{{\em JWST}\xspace}\xspace}
\newcommand{\hst}{{\em Hubble Space Telescope (HST)}\renewcommand{\hst}{{\em HST}\xspace}\xspace}
\newcommand{\ifss}{integral-field spectrographs (IFSs)\renewcommand{\ifss}{IFSs\xspace}\renewcommand{\ifs}{IFS\xspace}\xspace}
\newcommand{\ifs}{integral-field spectrograph (IFS)\renewcommand{\ifss}{IFSs\xspace}\renewcommand{\ifs}{IFS\xspace}\xspace}
\newcommand{\ifus}{integral field units (IFUs)\renewcommand{\ifus}{IFUs\xspace}\xspace}
\newcommand{\mos}{multi-object spectrograph (MOS)\renewcommand{\mos}{MOS\xspace}\xspace}
\newcommand{\goodss}{Great Observatories Origins Deep Survey (GOODS)-S\renewcommand{\goodss}{GOODS-S\xspace}\xspace}
\newcommand{\goods}{Great Observatories Origins Deep Survey (GOODS)\renewcommand{\goods}{GOODS\xspace}\xspace}
\newcommand{\scmos}{scientific CMOS (sCMOS)\renewcommand{\scmos}{sCMOS\xspace}\xspace}
\newcommand{\aof}{Adaptive Optics Facility (AOF)\renewcommand{\aof}{AOF\xspace}\xspace}
\newcommand{\dsp}{digital signal processor (DSP)\renewcommand{\dsp}{DSP\xspace}\renewcommand{\dsps}{DSPs\xspace}\xspace}
\newcommand{\dsps}{digital signal processors (DSPs)\renewcommand{\dsp}{DSP\xspace}\renewcommand{\dsps}{DSPs\xspace}\xspace}
\newcommand{\capi}{Coherent Accelerator Processor Interface (CAPI)\renewcommand{\capi}{CAPI\xspace}\xspace}

\usepackage{amssymb}

\title[SHS improvement using total variation minimisation]{Sensitivity improvements
  for Shack-Hartmann wavefront
  sensors using total variation minimisation}

\newcommand{\tvm}{total variation minimisation (TVM)\renewcommand{\tvm}{TVM\xspace}\renewcommand{\Tvm}{TVM\xspace}\xspace}
\newcommand{\Tvm}{Total variation minimisation (TVM)\renewcommand{\tvm}{TVM\xspace}\renewcommand{\Tvm}{TVM\xspace}\xspace}

\author[A.\ G.\ Basden et al.]{A.\ G.\ Basden$^{1}$\thanks{E-mail:
    a.g.basden@durham.ac.uk (AGB)}, T.\ J.\ Morris$^1$,
  D.\ Gratadour$^2$ and E.\ Gendron$^2$ \\
$^{1}$Department of Physics, South Road, Durham, DH1 3LE, UK\\
$^2$ Observatoire de Paris, LESIA, France}

\begin{document}
\maketitle

\begin{abstract}
We investigate the improvements in Shack-Hartmann wavefront sensor
image processing that can be realised using total variation
minimisation techniques to remove noise from these images.  We perform
Monte-Carlo simulation to demonstrate that at certain signal-to-noise
levels, sensitivity improvements of up to one astronomical magnitude
can be realised.  We also present on-sky measurements taken with the
CANARY adaptive optics system that demonstrate an improvement in
performance when this technique is employed, and show that this
algorithm can be implemented in a real-time control system.  We
conclude that total variation minimisation can lead to improvements in
sensitivity of up to one astronomical magnitude when used with
adaptive optics systems.
\end{abstract}


\section{Introduction}
\label{sect:intro}
Astronomical imaging on large telescopes is restricted in achievable
resolution by atmospheric turbulence which perturbs the wavefront of
incident light.  A solution to this problem is the use of \ao systems
\citep{adaptiveoptics} which use one or more wavefront sensors to
measure these perturbations, and a \dm to actively compensate them in
real-time.  The most commonly used wavefront sensor for astronomical
systems is a Shack-Hartmann sensor \citep{shs} which divides the
telescope pupil plane into an array of sub-apertures and is then used to
measure the incident instantaneous wavefront tilt across these.  

The isoplanatic patch of the atmosphere limits the \ao corrected
distance from the guide star on which the wavefront sensor is focused
to about 10~arcseconds, i.e.\ regions further from the guide star are
essentially uncorrected by the adaptive optics.  Additionally, since
the wavefront sensors require enough light to reliably detect target
motion on millisecond timescales within each sub-aperture, bright
targets are required.  Therefore, the fraction of the sky that is
observable with \ao correction, the sky-coverage, is typically only a
few percent for most \ao systems.  Improving sky-coverage for \ao
systems is therefore desirable, requiring increased sensitivity
wavefront sensors.

\Tvm \citep{rudin} is a process that attempts
to reduce the total variance within a signal.  Typically, a signal
containing significant noise will have a high total variation, and the
integrated absolute gradient will be large.  Minimisation of the total
variation of this signal, subject to the constraint that the cleaned
signal remains a close match to the original, removes unwanted noise,
while retaining important features such as sudden gradient changes.
\Tvm is very effective at removing noise in
flat image regions whilst also maintaining image features.
Alternative approaches such as median filtering and smoothing also
reduce the noise, but unfortunately remove features such as sharp
edges which may be inherent to the image, and therefore are not
appropriate where high image fidelity is necessary.

\begin{figure}
\includegraphics[width=0.3\linewidth]{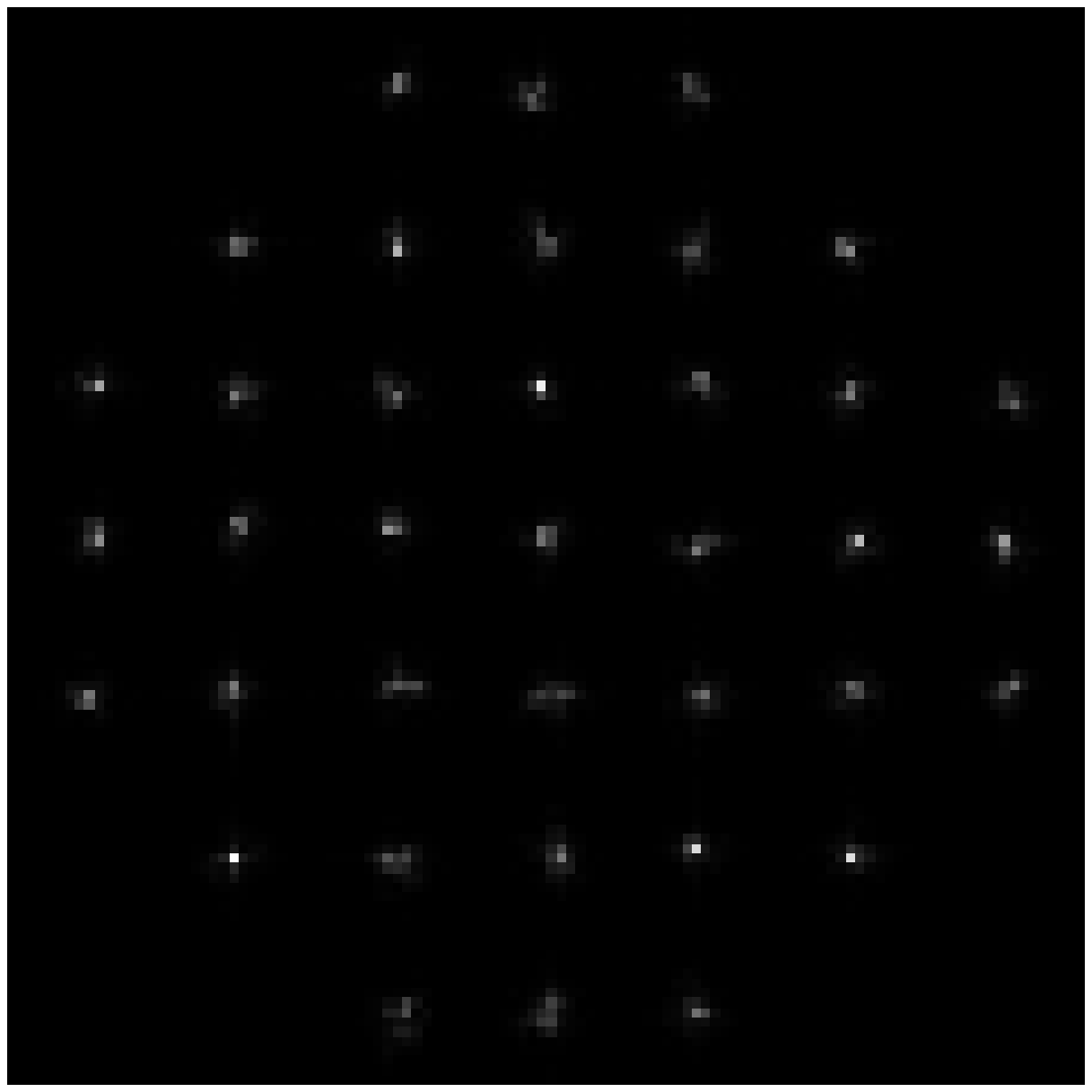}
\includegraphics[width=0.3\linewidth]{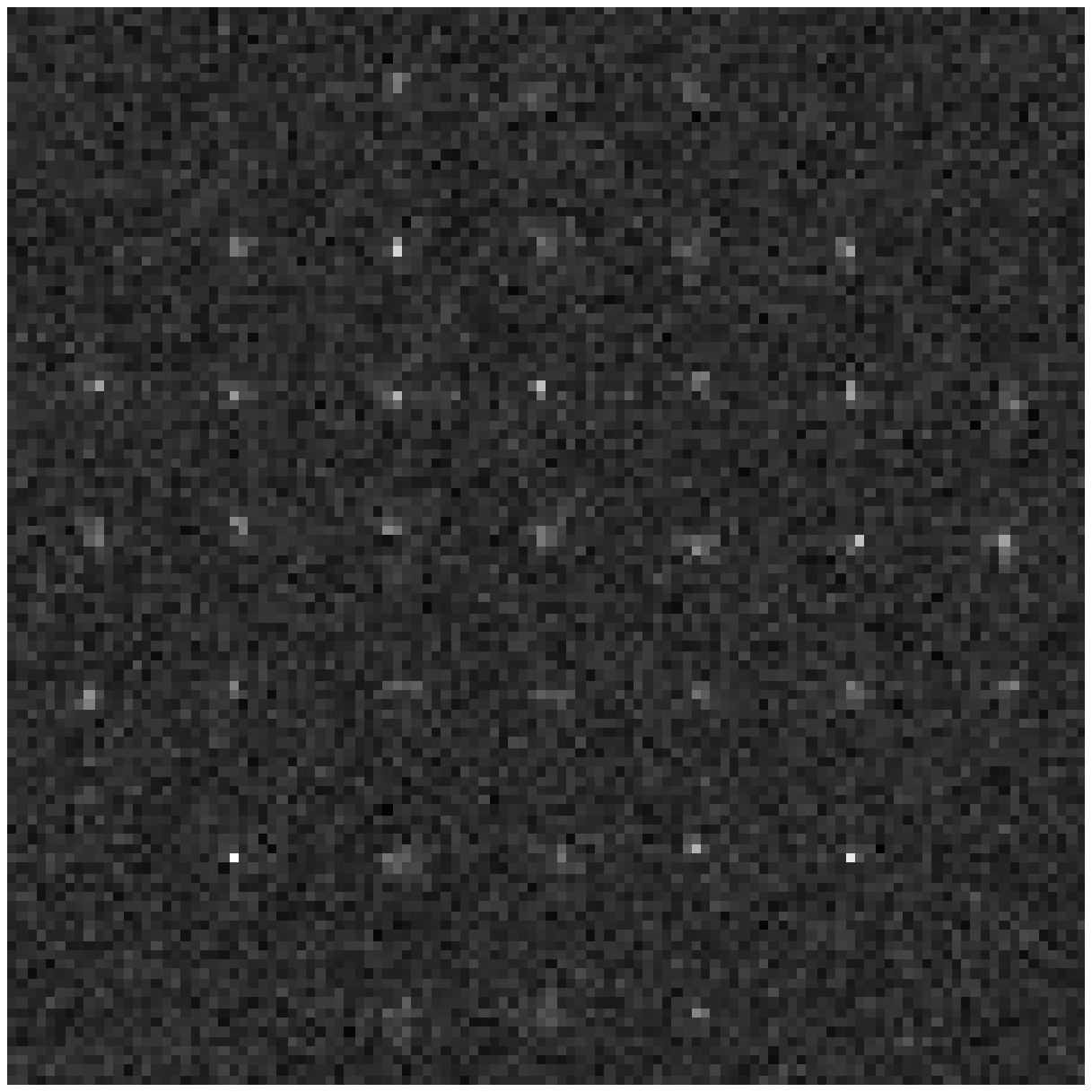}
\includegraphics[width=0.3\linewidth]{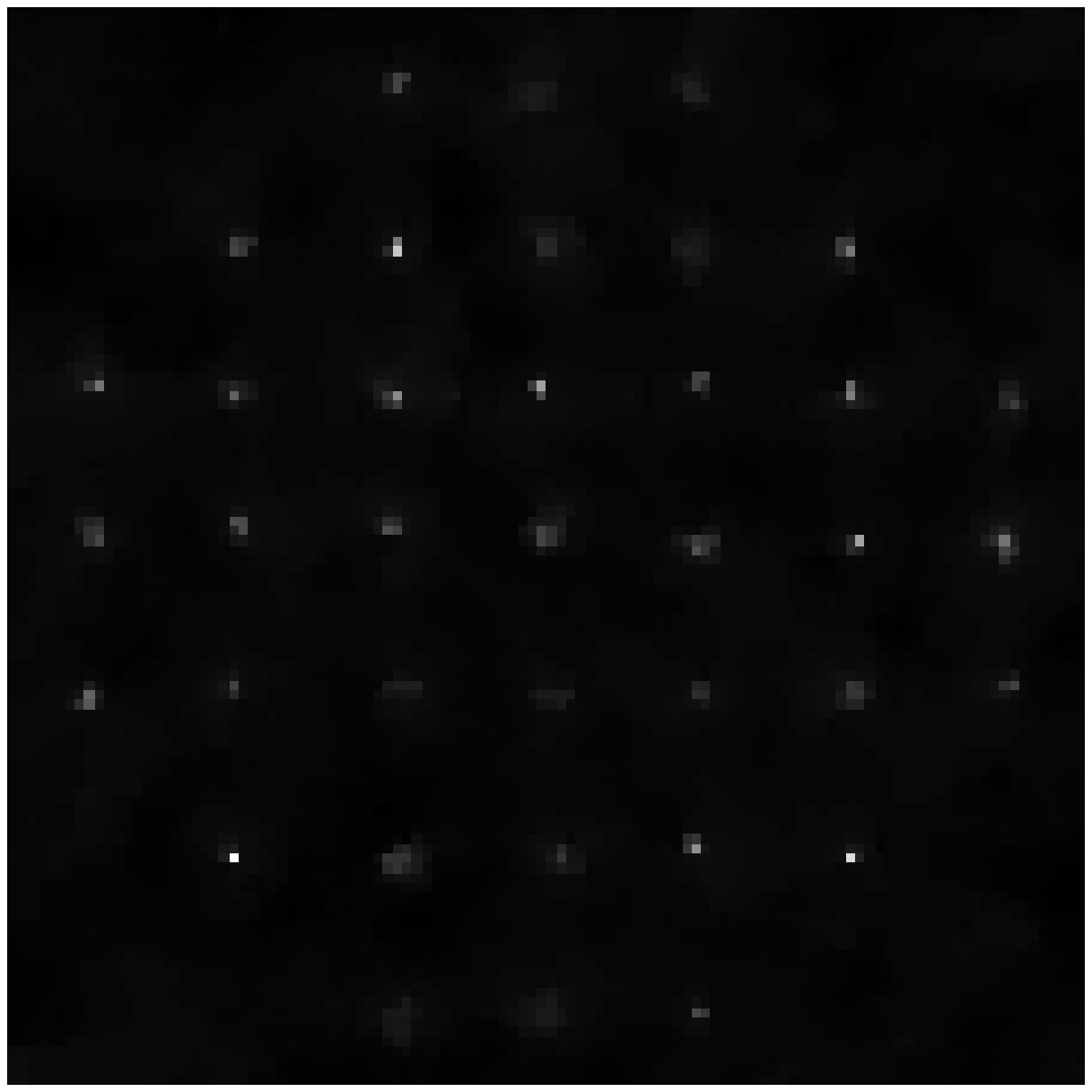}
\caption{(a) An example simulated noiseless Shack-Hartmann wavefront sensor
  image, (b) with photon and readout noise added, (c) after noise
  removal using TVM.}
\label{fig:shs}
\end{figure}

Fig.~\ref{fig:shs}(a) shows a noiseless Shack-Hartmann wavefront
sensor image.  Within this image, there are many sharp peaks, the
position of which (relative to some reference position) requires
determination with high precision, so that the corresponding incident
wavefront can be reconstructed.  Once noise is added
(Fig.~\ref{fig:shs}(b)), this determination becomes more difficult
with increased uncertainty.  By applying the principle of \tvm
 (Fig.~\ref{fig:shs}(c)), noise levels can be
reduced, resulting in improved spot position determination.

In \S2 we provide details of the \tvm
algorithm that we have investigated, and of the investigations that we
have performed.  In \S3 we present our results, including on-sky
measurements, and we conclude in \S4.

\section{Total variation minimisation model and application details}

We use a convergent algorithm developed by \citet{chambolle} for the
minimisation of total variation of an image.  This algorithm has
applications beyond image noise removal, for example image scaling; here
however we concentrate only on noise removal. 

It is assumed that an observed image, $g$ is the addition of an {\em a
  priori} piecewise smooth (or with little oscillation) image, $u$ and
a random Gaussian noise of estimated variance $\sigma^2$.  Therefore
the original image is estimated by solving \citep{chambolle}
\begin{equation}
\mathrm{min}\left( ||u-g||^2 = N^2\sigma^2 \right)
\end{equation}
where $N^2$ is the number of pixels.  

Throughout this paper, we consider noise reduction applied to
individual sub-apertures, rather than noise reduction applied to the
whole wavefront sensor image, because this is of most relevance to an
on-sky situation: within an \ao real-time control system, system
latency is reduced (and hence performance improved) if individual
wavefront sensor sub-apertures are processed separately, as soon as
the relevant pixel data arrive at the real-time computer, rather than
waiting and processing a whole frame at once.  The ability to access
the camera pixel stream depends somewhat on camera mode; however with
the CANARY \ao system \citep{canaryshort} and many others, camera
customisation, interface development, and custom software has made
this possible.  CANARY uses \darc for wavefront control
\citep{basden9,basden11} which is optimised for low latency operation,
and has the ability to process sub-apertures individually once pixels
become available.  We have therefore implemented the \tvm algorithm
within this system (which we discuss in \S\ref{sect:darc}), applying
\tvm on a per-sub-aperture basis.

\subsection{Monte-Carlo simulation of total variation minimisation performance}
There are many parameters that need to be considered when
investigating sub-aperture slope estimation improvement, including the
size of the spots (determined by optics and atmospheric conditions),
the signal level of the target, detector readout noise, and the number
of pixels within each sub-aperture.

We investigate performance of the noise removal algorithm spanning this
parameter space using Monte-Carlo simulation techniques.  Our
procedure is as follows:
\begin{enumerate}
\item A sub-aperture spot is generated at a random, known, position  ($S_\textrm{true}$).
\item Noise (photon and readout) is added.
\item Spot position is estimated using a centre of gravity algorithm  ($S_\textrm{sys}$).
\item \Tvm is applied to the image.
\item Spot position is estimated using a centre of gravity algorithm ($S_\textrm{tvm}$).
\item Steps 1--5 are repeated many ($N$) times.
\item The performance metric is calculated.
\end{enumerate}
The performance metric is given by
\begin{equation}
R = \frac{\sum_{m=1}^N
    |\left(S_\textrm{true}(m)-S(m)  \right)|}{N}
\label{eq:metric}
\end{equation}
where $N$ is the number of measurements taken and $S(m)$ is the
$m^{th}$ individual slope measurement measured with the $m^{th}$
Monte-Carlo realisation, either the true position, or the estimated
position (with noise added, $S_\textrm{sys}$, and after application of
noise removal, $S_\textrm{tvm}$).  In essence, the
absolute offset between estimated and true positions are computed, and
the mean offset calculated over ten thousand realisations.  We
refer to $R$ as the slope error, or slope estimation accuracy, and to
$S_\textrm{sys}$ as the ``No tvm'' case.

We consider signal levels from high light level, down to very low (10
photons per sub-aperture i.e.\ too low for good \ao correction, but
still of academic interest).  We consider a range of detector readout
noise from 0.1 to 16 electrons, which includes the parameter space for
the \emccds and \scmos detectors that are candidate wavefront sensors,
and also that corresponding to an electronically shuttered laser guide
star wavefront sensor that was used with CANARY.  Sub-aperture sizes
are considered from $8\times8$ to $16\times16$ pixels, corresponding
to the sizes used for CANARY, and also those that are likely for
wide-field \elt-scale \ao instruments.  Spot sizes are investigated
ranging from Nyquist sampled, to spots with a FWHM of about four
pixels, i.e.\ towards the practical upper size limit with which a
typical \ao system would work.

In conjunction with the noise removal algorithm under consideration
here, we use a background subtraction algorithm based on brightest
pixels \citep{basden10}, which sets the image background threshold
level (both noisy and denoised) at a level such that a given number of
image pixels remain above this threshold in each sub-aperture.  When
investigating performance, we use the number of retained image pixels
that gives best performance.

\subsection{On-sky testing of total variation minimisation}
\label{sect:darc}
We have implemented the \tvm algorithm within
the \darc system that is used by CANARY.  This is a dynamically
loadable modular control system, and so the introduction of new
algorithms does not require a modification of the core system, and
these algorithms can be loaded and unloaded from the real-time system
without affecting its subsequent operation, making it ideal for
algorithm development.

Our implementation includes three adjustable parameters, which can be
altered on a sub-aperture basis (allowing optimised operation with
wavefront sensors where the spot \psf varies across the sensor, for
example differing elongation when using laser guide stars).  These are
the ``strength'' (estimated noise standard deviation) of the noise
removal, the tolerance level (at which the image is considered
denoised), and the maximum number of iterations allowed (to avoid
significant increase to \ao system latency).  The maximum number of
iterations is set to a number greater than that typically required for
convergence (in which case, the algorithm does not perform all
iterations); it serves to prevent real-time system jitter in
rare cases where the algorithm is not converging quickly.

During these tests, we operate CANARY in a \scao mode, using a single
on-axis wavefront sensor, and interleave processing with and without
\tvm while measuring performance.  Strehl
ratio of the \ao corrected image is our performance metric, computed
on-axis using standard CANARY tools \citep{canaryresultsshort}.

\section{Discussion of performance improvements using TVM}
\begin{figure}
\includegraphics[width=0.45\linewidth]{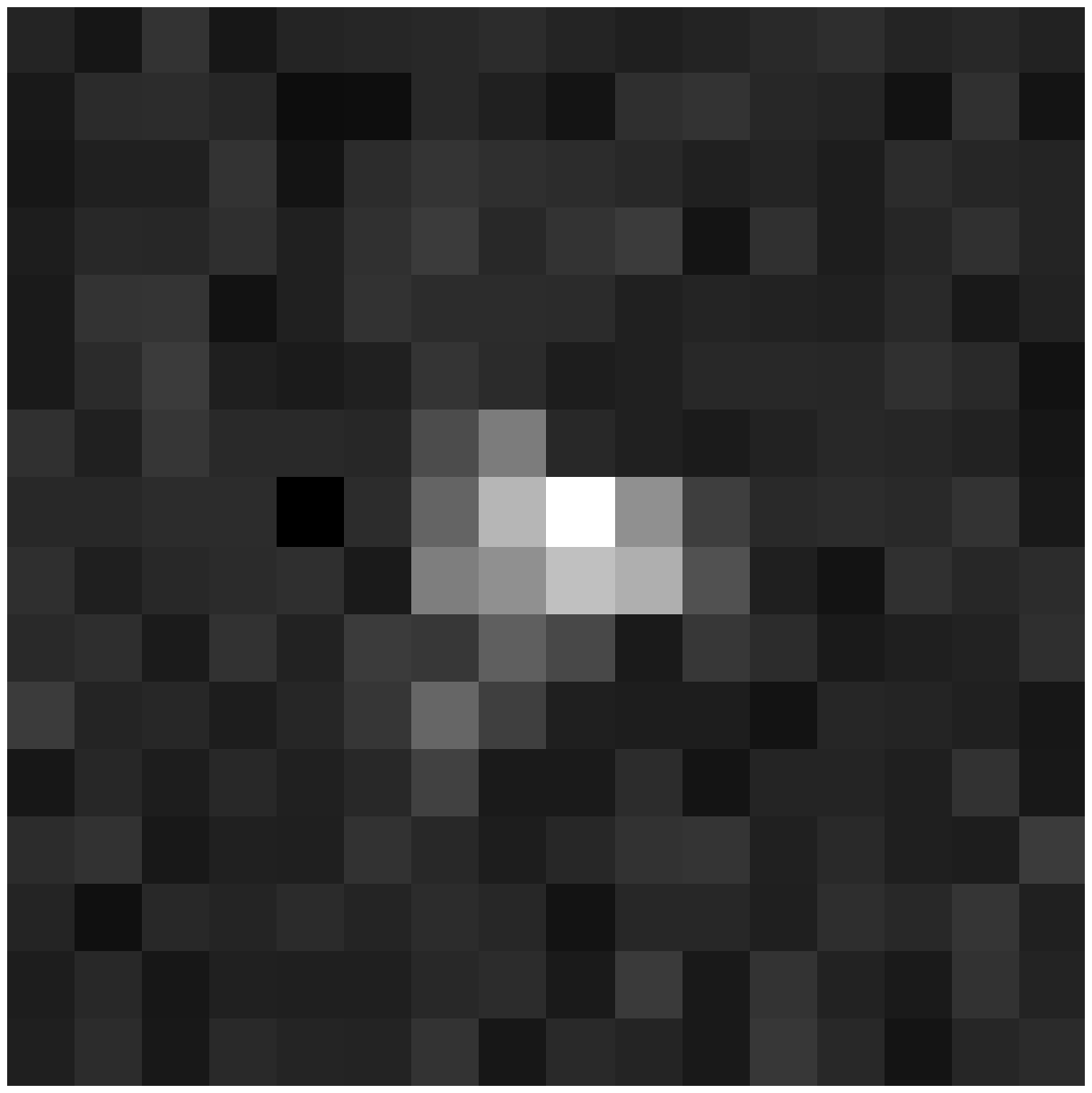}
\includegraphics[width=0.45\linewidth]{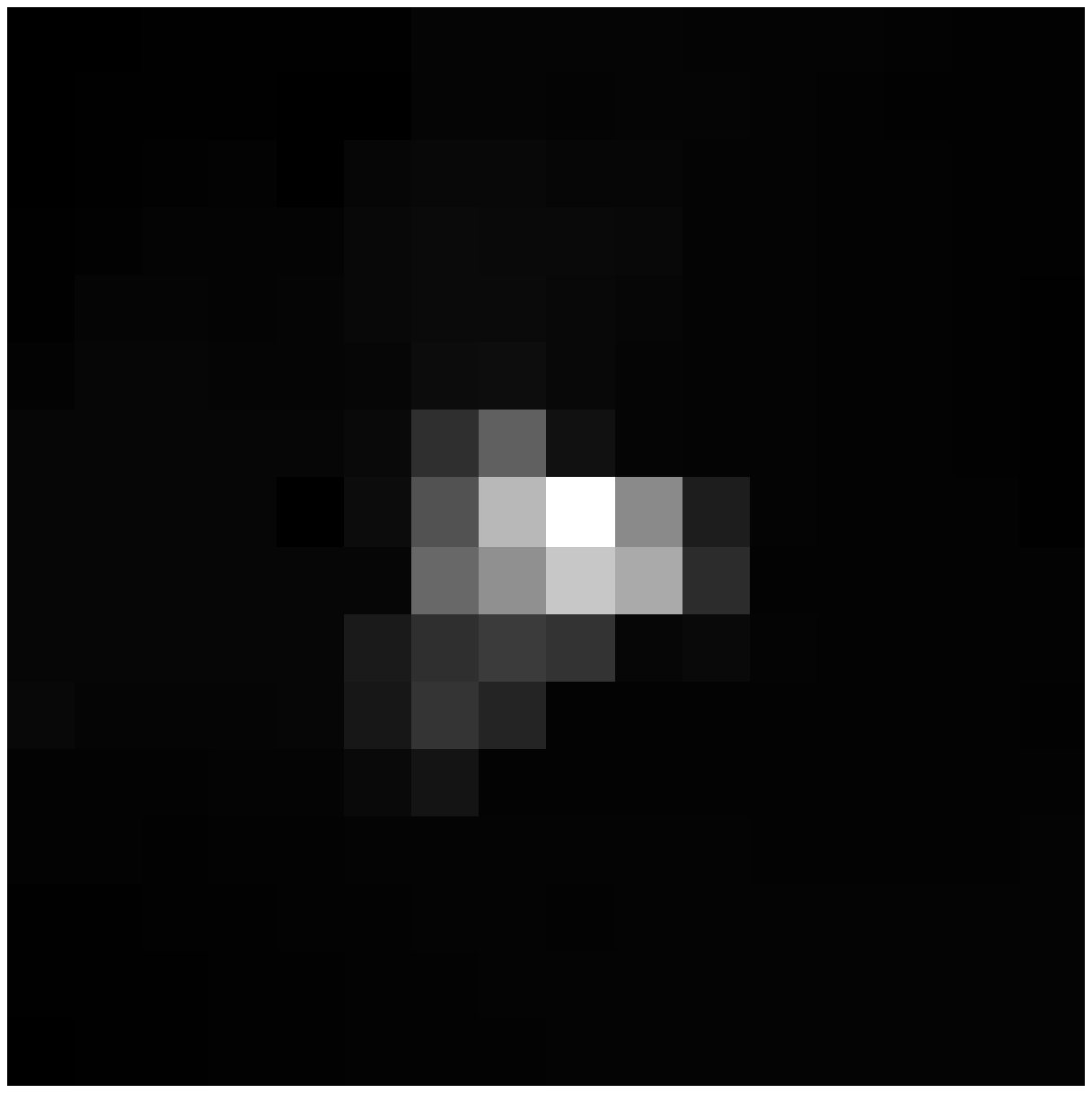}
\caption{A comparison showing a noisy Shack-Hartmann spot, and the
  same image after application of the TVM
  algorithm.}
\label{fig:shsspot}
\end{figure}

Accuracy of Shack-Hartmann slope estimation has been investigated in
simulation, comparing noisy and denoised Shack-Hartmann spots.
Fig.~\ref{fig:shsspot} shows a comparison of a simulated noisy and
denoised Shack-Hartmann spot.  In this figure, it is clearly evident
that the \tvm algorithm is effective at reducing the noise within this
image.

\subsection{Performance as a function of spot size}
\begin{figure}
\includegraphics[width=\linewidth]{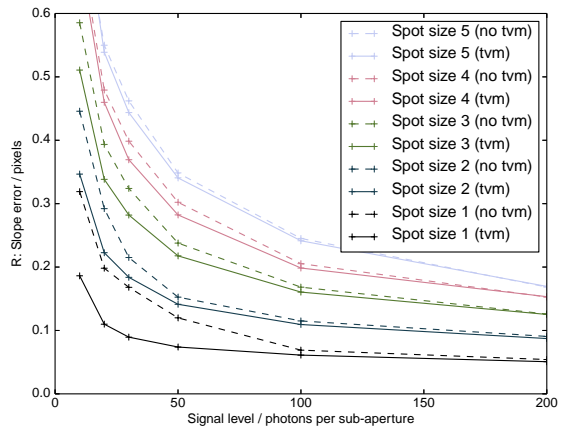}
\caption{A figure showing Shack-Hartmann slope estimation error as a
  function of signal level for cases with different Shack-Hartmann
  spot sizes, given as a function of Nyquist sampled size in the
  legend.  Solid lines show performance with TVM, while dashed lines
  are without.  Uncertainties are at the 1\% level and are not shown
  for clarity.}
\label{fig:spot}
\end{figure}

The size of a Shack-Hartmann spot determines, for a given flux level,
the intensity level of the brightest pixels, since incident flux is
spread over the spot.  Fig.~\ref{fig:spot} shows performance as a
function of signal level for different spot sizes, processed both with
and without \tvm.  Here, a $16\times16$ sub-aperture has been used,
with 0.1 electron readout noise.  From this figure, it is evident that
smaller spot sizes benefit most from \tvm, since the difference
between the noisy and denoised cases is largest.  When using \tvm, a
reduction in signal level by a factor of between 2--3 is possible
while still maintaining the non-\tvm performance level, i.e.\ guide
star magnitude can be decreased by up to one astronomical order of
magnitude when using \tvm.

\subsection{Performance as a function of sub-aperture size}
\begin{figure}
\includegraphics[width=\linewidth]{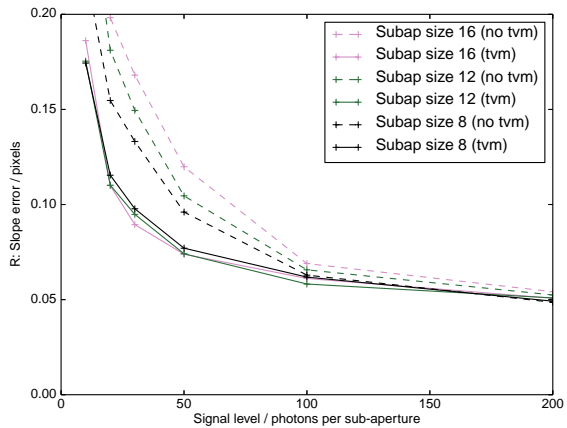}
\caption{A figure showing Shack-Hartmann slope estimation error as a
  function of signal level for different sized sub-apertures, with the
  linear dimension given in the legend.  Solid lines show performance
  with TVM, while dashed lines are without.  Uncertainties are at the
  1\% level and are not shown for clarity.}
\label{fig:nx}
\end{figure}

For a given spot size, a larger sub-aperture will contain a larger
number of pixels with just noise (i.e.\ negligible useful signal).
However, in some cases, a large sub-aperture may be necessary, for
example in open-loop \ao systems, where a wide-field of view is
required to detect large spot motions.  Fig.~\ref{fig:nx} shows slope
estimation error as a function of signal level, for different
sub-aperture sizes.  Here it is interesting to note that in the
denoised case (with \tvm), performance is essentially unrelated to
sub-aperture size, since the \tvm is successfully removing the
background noise.  However, in the noisy cases, performance gets worse
as sub-aperture size increases as expected due to the presence of an
increased number of noisy pixels.  In this figure, readout noise is
set at 0.1 electrons, and the Shack-Hartmann spot is Nyquist sampled.

\subsection{Performance at low signal-to-noise ratios}
\begin{figure}
\includegraphics[width=\linewidth]{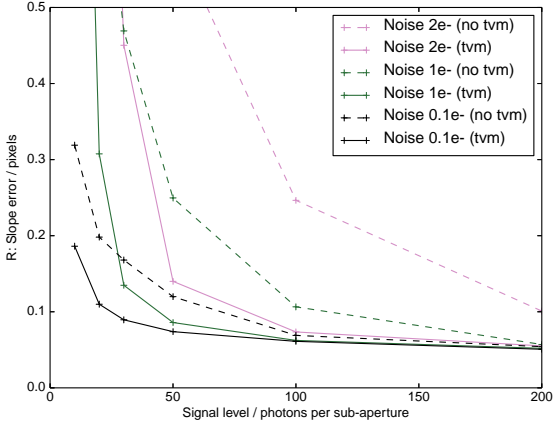}
\caption{A figure showing Shack-Hartmann slope estimation error as a
  function of signal level for cases with different readout noise (as
  given in the legend).  Solid lines show performance with TVM, while
  dashed lines are without.  Uncertainties are at the 1\% level and
  are not shown for clarity.}
\label{fig:noise}
\end{figure}

Fig.~\ref{fig:noise} shows performance as a function of signal level
for different wavefront sensor readout noise, for a $16\times16$ pixel
sub-aperture with a Nyquist sampled spot.  Here it can be seen that as
signal level is reduced, the slope error ($R$) in cases without \tvm
increases faster than with.  At certain signal levels, using \tvm
allows operation at light levels a factor of 2--3 times lower than
without \tvm, whilst achieving the same slope estimation accuracy.
For example, with a readout noise of 0.1 electrons, a signal level of
30 photons with \tvm gives the same performance as 80 photons without \tvm.

\subsection{Discussion of background level selection using brightest
  pixels}
\begin{figure}
\includegraphics[width=\linewidth]{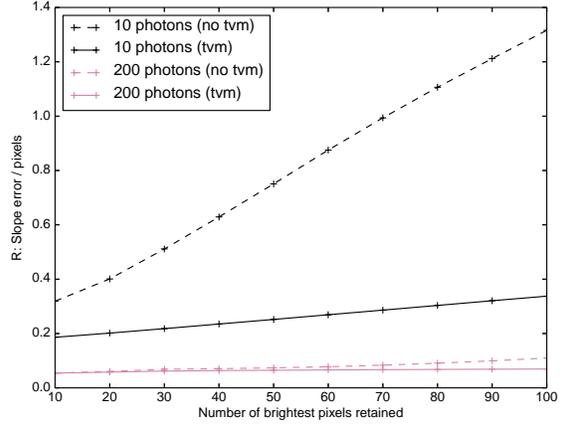}
\caption{A figure showing slope error as a function of number of
  brightest pixels retained after background level removal prior to
  slope estimation.  Results for two different light levels are shown,
  with and without TVM, as noted in the legend.  Uncertainties are at
  the 1\% level and are not shown for clarity.}
\label{fig:ubsig}
\end{figure}

\begin{figure}
\includegraphics[width=\linewidth]{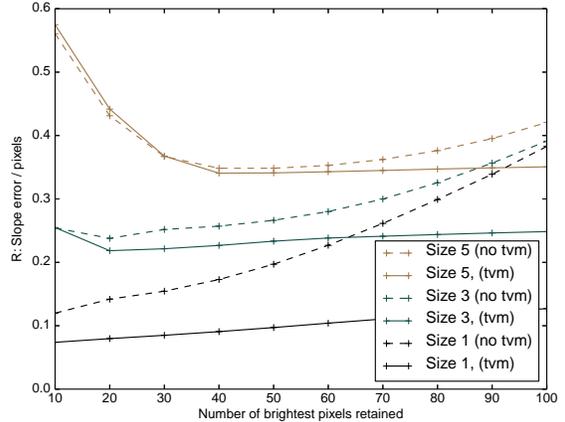}
\caption{A figure showing slope error as a function of number of
  brightest pixels retained after background level removal prior to
  slope estimation.  Results for different sub-aperture PSF spot sizes
  are shown, with and without TVM, as noted in the legend, where the
  size is the PSF oversampling factor beyond Nyquist sampled.
  Uncertainties are at the 1\% level and are not shown for clarity.}
\label{fig:ubsize}
\end{figure}

\begin{figure}
\includegraphics[width=\linewidth]{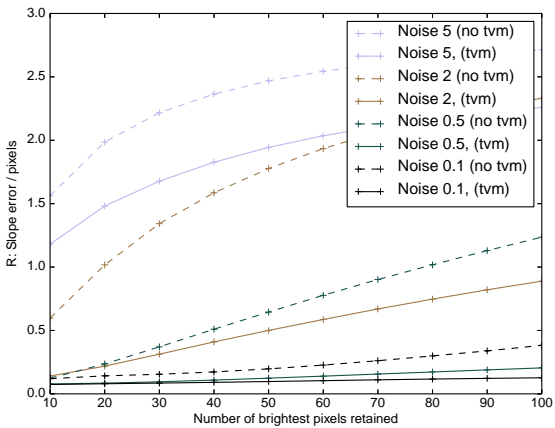}
\caption{A figure showing slope error as a function of number of
  brightest pixels retained after background level removal prior to
  slope estimation.  Results for different readout noise levels (in
  electrons) are shown, with and without TVM, as noted in the legend.
  Uncertainties are at the 1\% level and are not shown for clarity.}
\label{fig:ubnoise}
\end{figure}

So far, we have been using the number of pixels for background
selection that give best performance, both for the noisy and denoised
cases.  However, it is useful to investigate how this background affects
performance.  Fig.~\ref{fig:ubsig} shows slope error ($R$) as a function of
number of brightest pixels retained for two different signal levels
(assuming 0.1 electrons readout noise, a $16\times16$ sub-aperture and
a Nyquist sampled spot).  It is evident here that when there are fewer
photons, \tvm is effective at removing the effect of noise, so that
the final slope error ($R$) is less dependent on the number of brightest pixels
retained.  Fig.~\ref{fig:ubsize} shows slope error ($R$) as a function of
number of brightest pixels retained for different spot sizes.  Here,
it is clear that again, \tvm removes some of the sensitivity to
background level, since noise has effectively been removed (a signal
level of 50 photons is assumed).  Similarly, using the above
assumptions, Fig.~\ref{fig:ubnoise} shows slope error ($R$) at different
detector readout noise levels, again displaying the improvements
brought by the \tvm algorithm.  

\ignore{Looking at py/tvm/plotUB.py...}

When selecting the number of brightest pixels to retain during
background level thresholding, the main consideration should be given
to the size of the sub-aperture \psf.  Using \tvm provides a key
benefit of reducing the dependency on accurate background subtraction.  

\subsection{Application to laser guide star elongated spots}
\begin{figure}
\includegraphics[width=\linewidth]{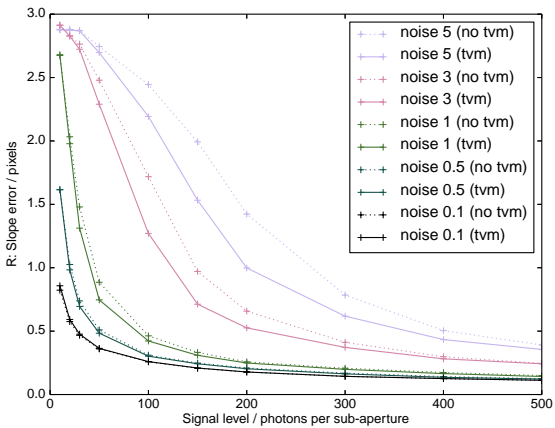}
\caption{A figure showing predicted Shack-Hartmann slope error as a
  function of signal level for a $16\times16$ pixel sub-aperture with
  an elongated spot PSF ($3\times$ longer than wide).  Cases with and
  without TVM are given in the legend, for different detector readout
  noise.  Uncertainties are at the 1\% level and are not shown
  for clarity.}
\label{fig:lgs}
\end{figure}

We have also investigated the application of \tvm to elongated
Shack-Hartmann spots.  For the results presented here, we assume a
spot that is elongated by a factor of three, i.e. three times longer
in one dimension than the other.  Fig.~\ref{fig:lgs} shows performance
(slope error, $R$) as a function of signal level, for different readout
noise values, both with and without \tvm.  It can be seen here that
the benefit obtained from \tvm increases with readout noise, and
performance is never worse than without \tvm.  The performance
improvements are less marked than for the natural guide star case,
though the use of \tvm can enable the same slope prediction
performance for light levels reduced by up to about 25\% for the high
readout noise case with greater than three electrons.

\subsection{On-sky measurements}
\begin{figure}
\includegraphics[width=0.3\linewidth]{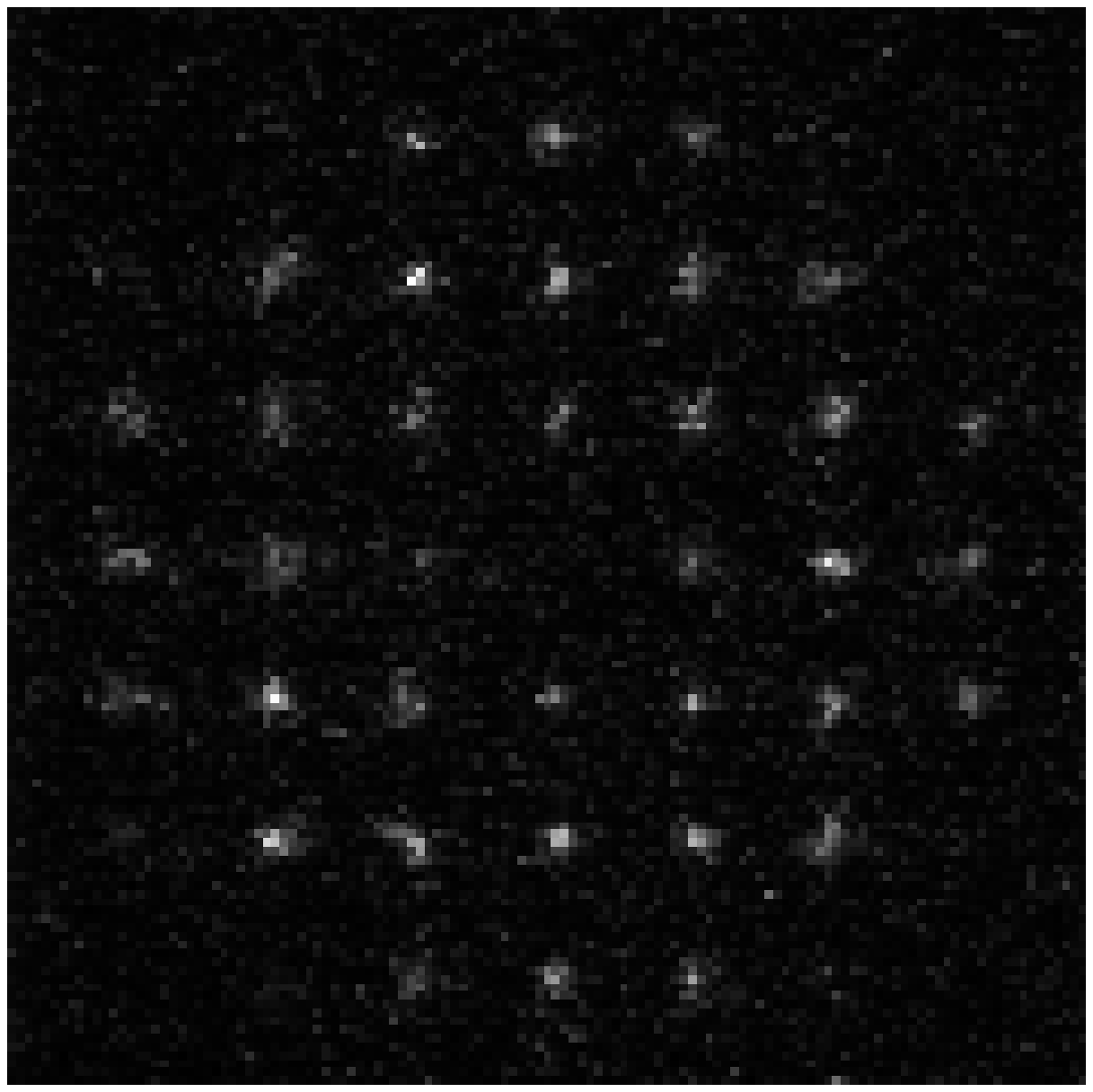}
\includegraphics[width=0.3\linewidth]{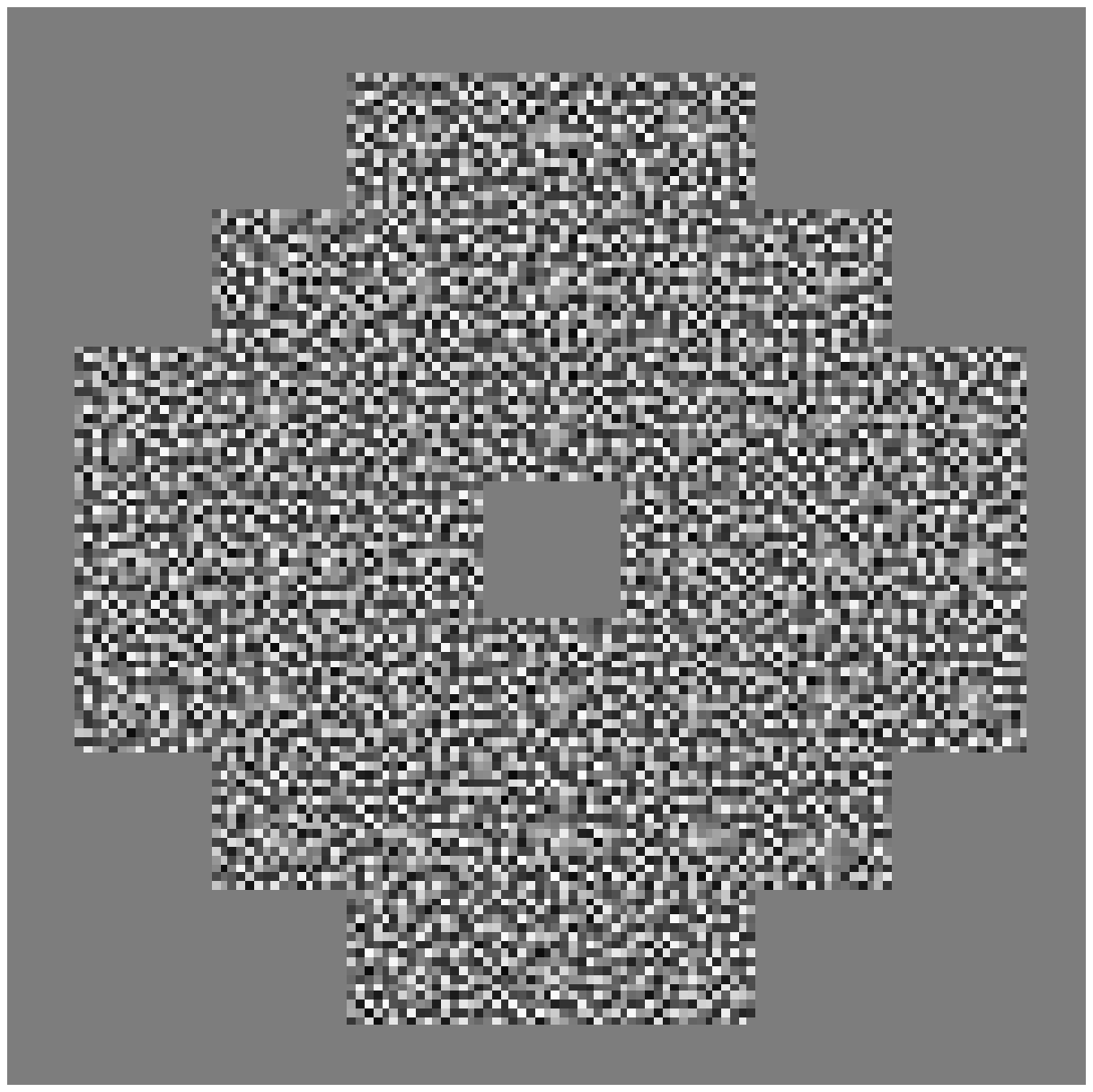}
\includegraphics[width=0.3\linewidth]{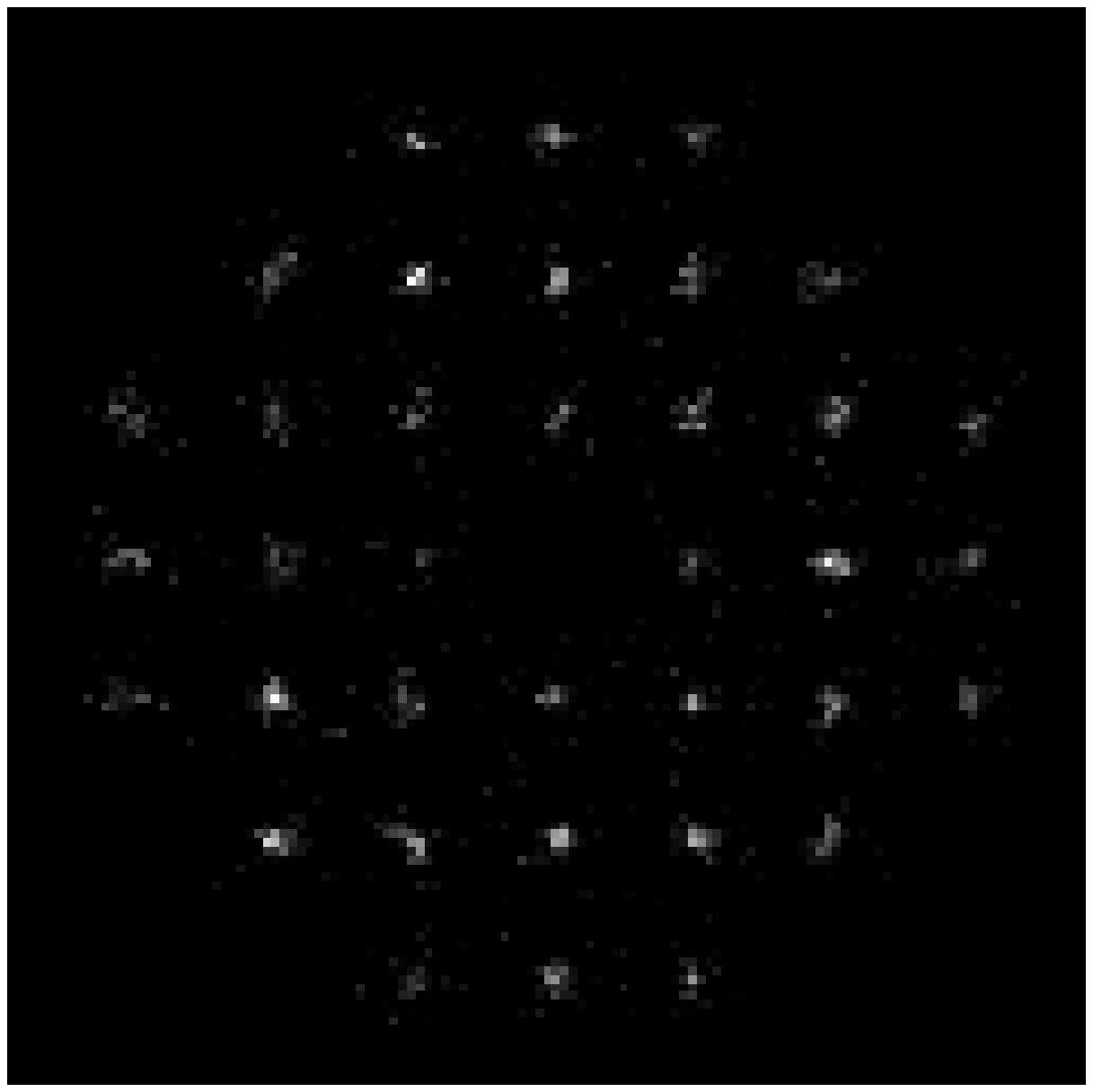}
\caption{(a) A single frame of a wavefront sensor image obtained on-sky by
  the CANARY AO system at the William Herschel Telescope.  (b) The
  noise removed using TVM with a variance of 1 (in active sub-apertures).  (c) Calibrated
  image after TVM and background subtraction.}
\label{fig:shsonsky}
\end{figure}

Because on-sky time was limited, we did not attempt to explore a
large parameter space of seeing conditions, spot size, signal level
and readout noise.  Rather, we have selected a target where noise is
evident within the sub-apertures (Fig.~\ref{fig:shsonsky}), and
operated the CANARY \ao system in \scao mode both with and without
\tvm.  The data presented here were taken on the night of 12th July
2014 with CANARY on the William Herschel Telescope, for just over one
hour from about 3am.

We have taken four sets of observations, during which the CANARY \scao
loop was closed, and H-band science images obtained.  Each set of
observations commenced with a measurement without \tvm, and then a
number of measurements (8 or 20) with different \tvm strength
(estimated noise standard deviation) factors (increasing monotonically
from 0 to 2).  Within the observation set, this was then repeated.
The detector used was an \emccd, and multiplication gain was set to
maximum for the first two observation sets, and 75\% for the last two,
allowing the signal level to be reduced.  Signal level was between
500--1000 detector counts.

\begin{figure}
\includegraphics[width=\linewidth]{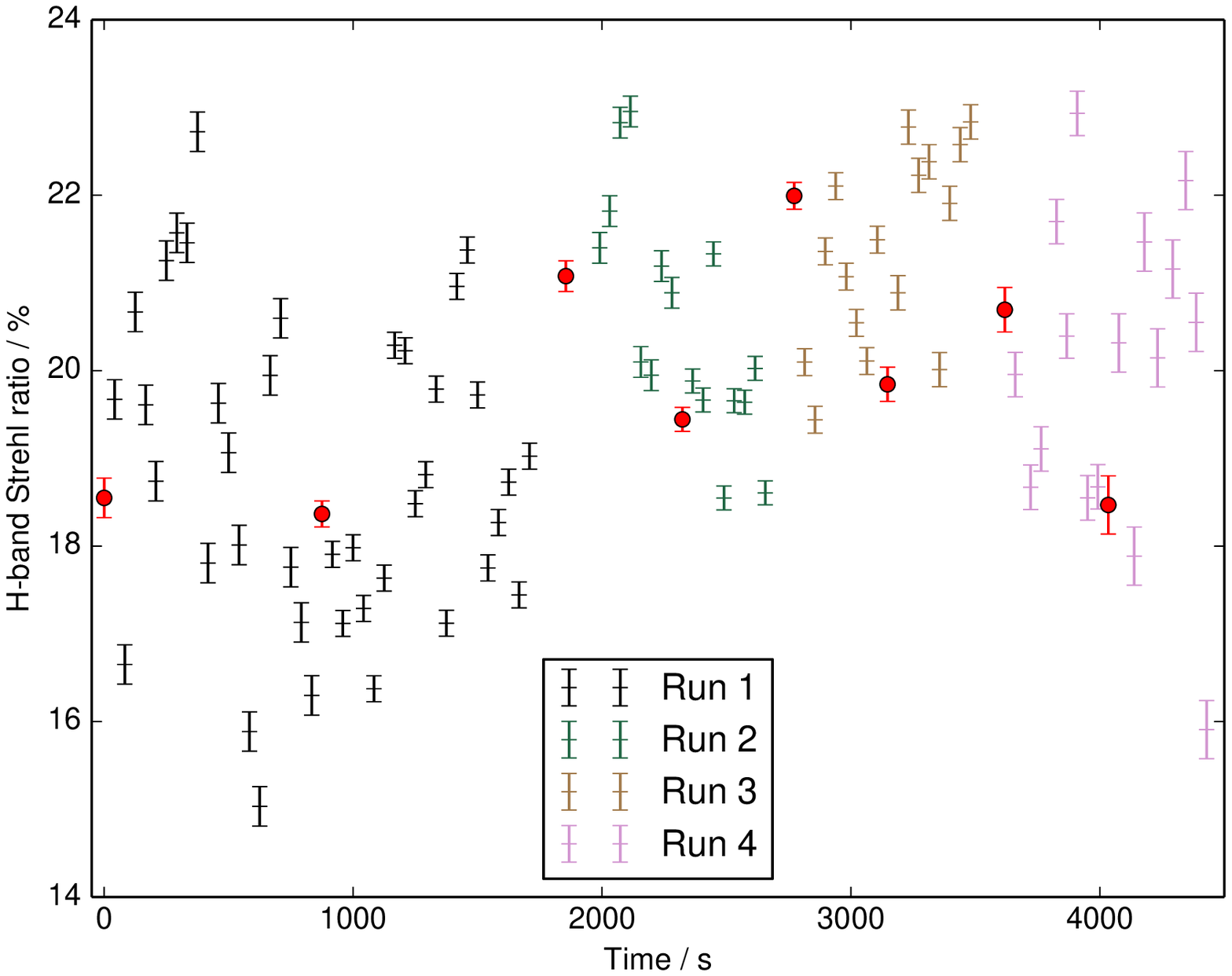}
\includegraphics[width=\linewidth]{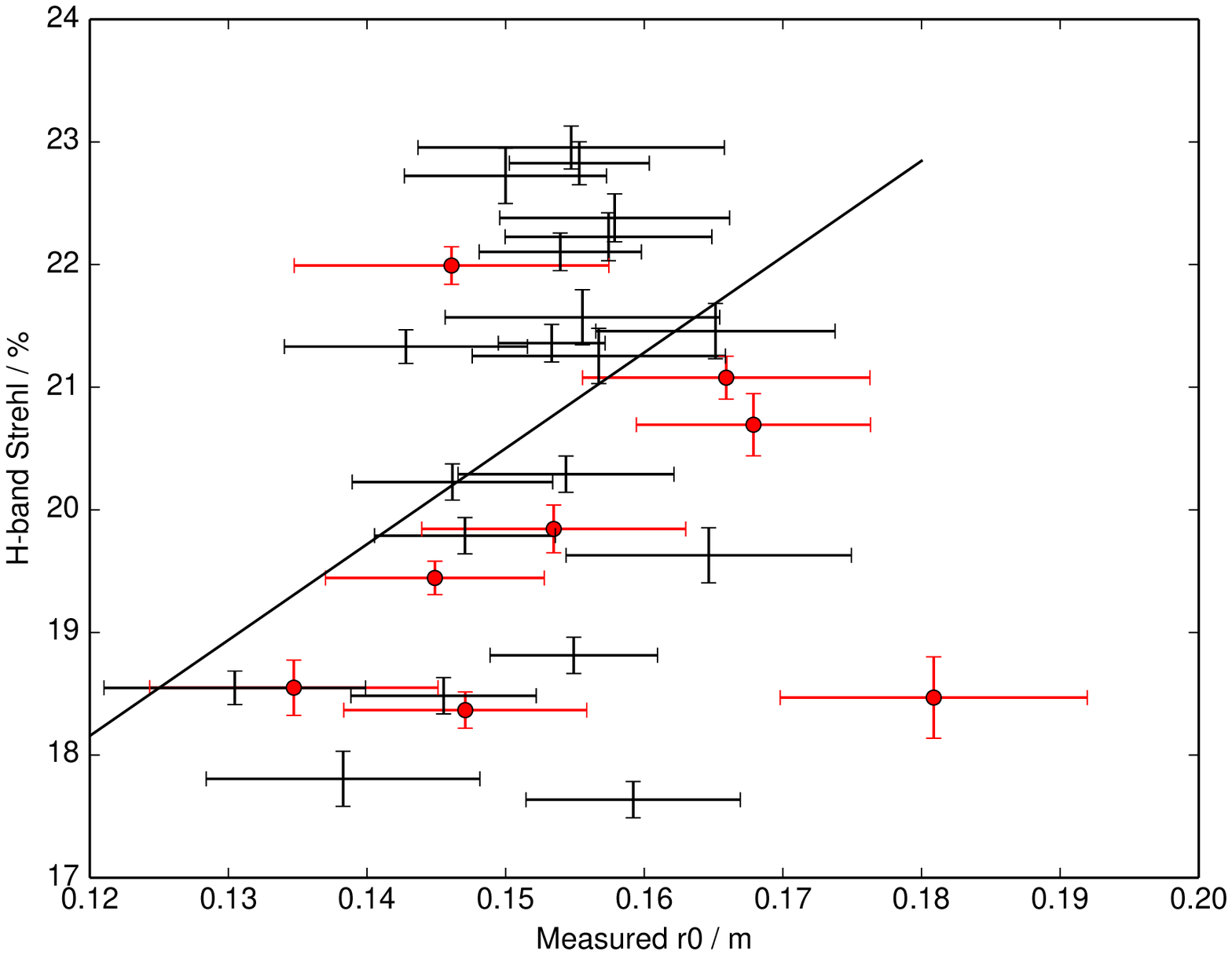}
\includegraphics[width=\linewidth]{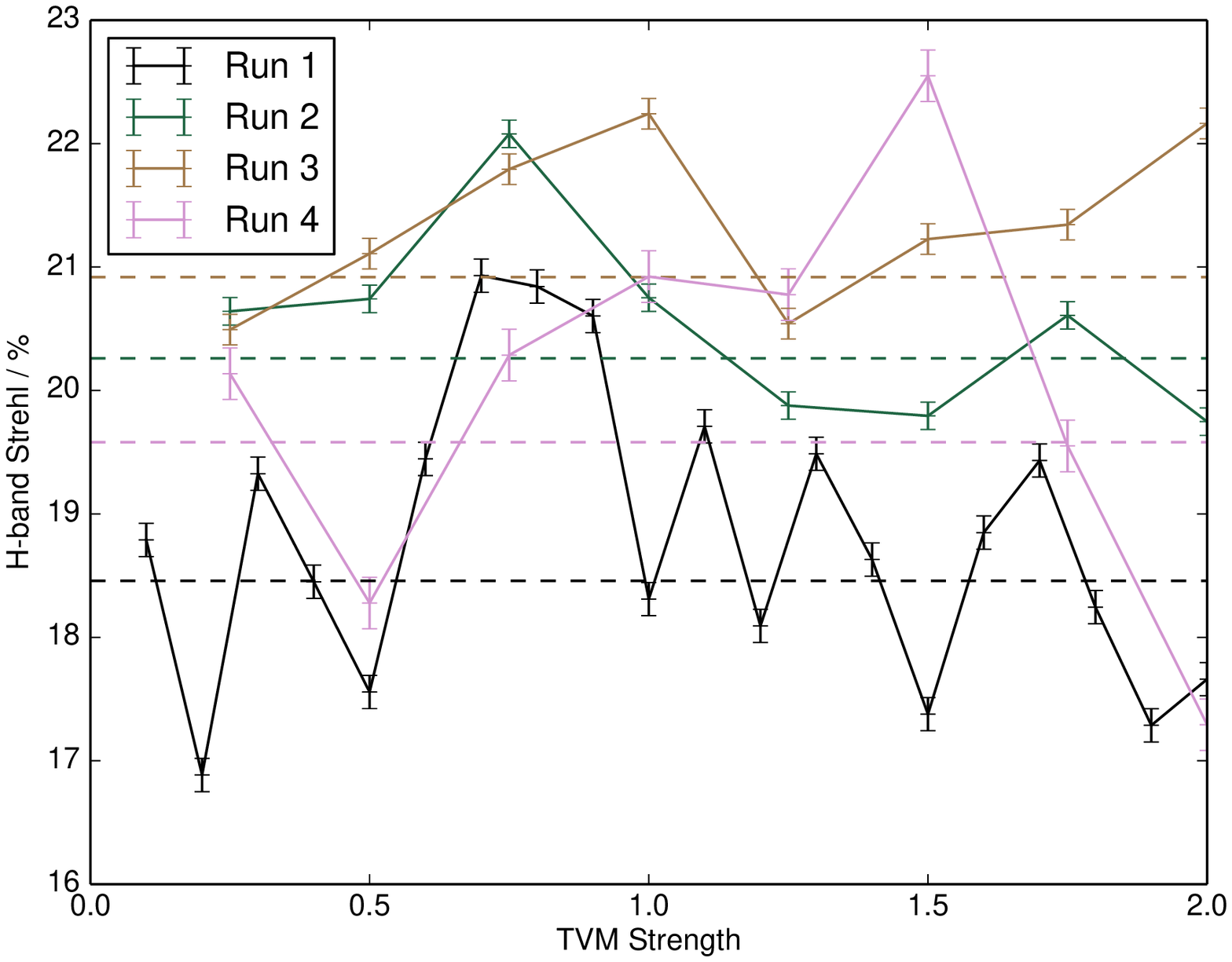}
\caption{(a) A figure showing H-band Strehl ratio obtained with CANARY
  operating in SCAO mode, both with and without TVM as a function of
  time from first observation.  The red circles show Strehl obtained
  without TVM, while other points are with TVM at different strengths.
  (b) H-band Strehl ratio as a function of $r_0$, with red circles
  representing measurements without TVM, and black crosses representing
  measurements with TVM.  (c) H-band Strehl ratio as a function of TVM
  noise standard deviation estimate.  The horizontal lines show Strehl
  obtained without TVM. }
\label{fig:tvmonsky}
\end{figure}

Results are shown in Fig.~\ref{fig:tvmonsky}, and provide
evidence that this technique is able to improve \ao system
performance, though, due to the lack of on-sky time, this is not
wholly conclusive.  Mean Strehl ratios for each observation set are shown in
table~\ref{tab:meanstrehl}, and in each case show an improvement when
using \tvm.  There is a large variation in performance as a function
of time, which is typical of the time-varying seeing conditions
commonly seen with CANARY.  It should be noted that the improvement
obtained using \tvm is, in this case, small:  it is likely that larger
improvements would be seen at other signal-to-noise regimes, though
this was not investigated on-sky.  

Because of the natural variability of seeing, we also show, in
Fig.~\ref{fig:tvmonsky}(b), the H-band Strehl ratio as a function of
Fried's parameter, $r_0$.  In this case, $r_0$ is computed from
reconstructed pseudo-open-loop slope measurements, which are, in turn
computed from a time-series of the on-axis closed-loop wavefront slope
measurements and the recorded \dm actuator commands.  The line fitted
through these data points is the best-fit to the cases with \tvm
implemented (with a regression correlation 0.38).  There is evidence
here that the \tvm performance is above the level of performance
without \tvm (represented by circles), indicating that the use of \tvm
has improved performance.  However, again, there is some uncertainty,
due to the lack of on-sky measurements, and so this should not be
taken as conclusive: the statistical significance is low.

There is some evidence from Fig.~\ref{fig:tvmonsky}(c)
that using estimated noise standard deviation of around 0.8 in the
\tvm gives best performance (though we acknowledge that this could
just be an artefact of the changing seeing conditions).  Therefore,
table~\ref{tab:meanstrehl} also includes the mean Strehl ratios
obtained by considering only noise standard deviation estimates (\tvm
strength) of between 0.6 to 1.1, and shows that indeed, using \tvm in
this regime leads to a further increase in \ao performance.

\begin{table}
\centering
\begin{minipage}{7cm}
\caption{AO system performance as given by H-band Strehl ratio (\%)
  for CANARY operating in SCAO mode, both without and with TVM (with
  results averaged over all strengths, or estimated noise standard deviation,
  investigated) during the four different observations made (Obs).
  Also shown is performance using TVM with strength restricted to
  between 0.6--1.1.}
\label{tab:meanstrehl}
\begin{tabular}{llll}\hline
Obs & No TVM & TVM & TVM strength 0.6--1.1\\ \hline
1 & $18.4 \pm 0.1 $ & $ 18.8 \pm 1.8 $ & $ 20.0 \pm 1.6 $\\
2 & $20.3 \pm 1.2 $ & $ 20.5 \pm 1.3 $ & $ 21.4 \pm 2.0 $\\
3 & $20.9 \pm 1.5 $ & $ 21.4 \pm 1.1 $ & $ 22.0 \pm 0.4 $\\
4 & $19.6 \pm 1.6 $ & $ 20.0 \pm 1.8 $ & $ 20.6 \pm 1.2 $\\ \hline
\end{tabular}
\end{minipage}
\end{table}

\subsection{High noise wavefront sensor cameras}
The wavefront sensors commonly used for laboratory \ao systems
typically have higher readout noise than those used on-sky, primarily
for reasons of cost.  We therefore investigate the performance of \tvm
for a sensor with a readout noise of 16 electrons, corresponding to
the Imperx Bobcat model that we use with the DRAGON \ao test-bench
\citep{dragon}.  Incidentally, this sensor was also briefly used
on-sky with CANARY as a substitute \lgs wavefront sensor (it has an
electronic shuttering capability) after a fault developed in the
previous sensor.

\begin{figure}
\includegraphics[width=\linewidth]{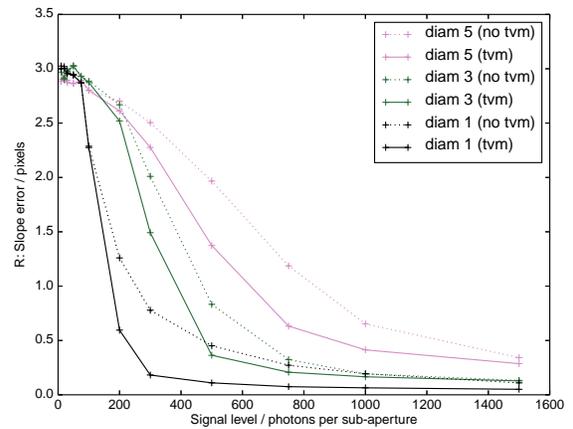}
\caption{A figure showing Shack-Hartmann slope error as a function of
  signal level for a $16\times16$ pixel sub-aperture and a detector
  with 16 electrons readout noise.  Lines for different PSF sizes are
  shown (given in the legend), with the numbers corresponding to the
  over-sampling factor (beyond Nyquist).  Cases with and without TVM
  are given in the legend.  Uncertainties are at the 1\% level and are
  not shown for clarity.}
\label{fig:bobcat}
\end{figure}

Fig.~\ref{fig:bobcat} shows slope error ($R$) as a function of incident
signal for different sub-aperture spot sizes.  As previously, the \tvm
algorithm provides an advantage for slope estimation, and gains up to
one astronomical magnitude in performance for this sensor.

\section{Conclusions}
We have investigated the use of a \tvm algorithm to improve slope
estimation accuracy with Shack-Hartmann wavefront sensor images.  We
find that in certain situations with low signal-to-noise ratio (with
appropriate signal and noise levels), the performance improvements
obtained can be equivalent to gaining an astronomical magnitude in
photon flux.  The use of \tvm never leads to a reduction in slope
estimation accuracy on average.  Larger sub-aperture sizes see most
benefit and so this is particularly relevant for open-loop \ao
systems.  Our investigation has shown that \tvm is applicable for both
\ngs and elongated \lgs Shack-Hartmann spots.  We have also presented
on-sky results from the CANARY \ao demonstrator instrument, which
provide evidence for successful improvement of on-sky \ao performance
using \tvm.

\section*{Acknowledgements}
This work is funded by the UK Science and Technology Facilities
Council, grant ST/I002871/1.  The author thanks the CANARY team for
allowing on-sky testing.

\bibliographystyle{mn2e}

\bibliography{mybib}
\bsp

\end{document}